\documentclass[pra,twocolumn,showpacs,superscriptaddress]{revtex4-2}
\usepackage{amsmath,amscd,amssymb,color,amsthm}
\usepackage{graphicx,amsfonts,dsfont}
\usepackage{epstopdf}
\usepackage{float}
\usepackage{hyperref}
\usepackage{enumerate,bbold}
\usepackage{array, comment, bm, braket}

\usepackage{color}
\usepackage{xcolor}
\definecolor{green1}{RGB}{10,150,0}
\usepackage{breakurl}
\begin{document}
\title{Neural networks learn to reconstruct multipartite
entanglement from quantum marginals}
\author{Matreyee Kandpal}
\email{ph23022@iisermohali.ac.in}
\affiliation{Department of Physical Sciences, Indian
Institute of Science Education \&
Research Mohali, Sector 81 SAS Nagar,
Manauli PO 140306 Punjab, India.}
\author{Arvind}
\email{arvind@iisermohali.ac.in}
\affiliation{Department of Physical Sciences, Indian
Institute of Science Education \&
Research Mohali, Sector 81 SAS Nagar,
Manauli PO 140306 Punjab, India.}
\author{Kavita Dorai}
\email{kavita@iisermohali.ac.in}
\affiliation{Department of Physical Sciences, Indian
Institute of Science Education \&
Research Mohali, Sector 81 SAS Nagar,
Manauli PO 140306 Punjab, India.}
\begin{abstract}
Different sets of local correlations are not equivalent: some
fragments of reduced data uniquely determine a global quantum
state, while others leave it ambiguous. 
The quantum marginal problem asks whether a collection of reduced density
matrices uniquely determines a compatible global quantum state.
Although generic quantum states are uniquely specified by suitable sets
of marginals, different collections of marginals are not equally
informative: some uniquely determine the global state, whereas others
leave it ambiguous. Identifying when uniqueness holds, and
reconstructing the global state from partial information, remains
computationally demanding and experimentally challenging.

We show that information about the multipartite
entanglement class and reconstructability in four-qubit systems
is compactly encoded in small sets of two- and three-qubit
marginals.  Using semidefinite programming, we chart the
reconstructability landscape across 49 inequivalent SLOCC
entanglement classes and show that uniqueness strongly depends
on both entanglement structure and marginal order.  Neural
networks trained only on reduced density matrices learn this
structure directly. They accurately classify marginal
reconstructability and, when uniqueness holds, reconstruct the
full four-qubit density matrix with high fidelity from two- and
three-qubit marginals.  
		
We benchmark the approach on a four-qubit nuclear magnetic
resonance quantum processor and demonstrate that
reconstructions from experimentally measured marginals remain
faithful despite phase damping and control imperfections.  Our
results show that neural networks can learn when local
correlations uniquely specify a global quantum state, and reveal how global
quantum structure is encoded in reduced data.  
\end{abstract} 

\maketitle 
%%%%%%%%%%%%%%%%%%%%%%%%%%%%%%%%%%%%%%%%%%%%%%%%%%%%%%%%%%%%%

\section{Introduction} 
\label{intro} 
In realistic quantum devices,
full access to the underlying quantum state is rarely available, and
experiments typically probe only partial information~\cite{cotler}. The
exponential growth of Hilbert space makes full quantum state tomography rapidly
infeasible, motivating the search for reduced descriptions that preserve
essential global state structure~\cite{Cramer2010}.  The quantum marginal
problem(QMP) asks  whether a given collection of reduced density matrices(RDMs)
of a multipartite system is compatible with a global quantum state, and if so,
whether that state is uniquely determined by the
marginals~\cite{klyachko2006}.  Beyond its computational complexity, the QMP
unearths fundamental constraints linking local reduced data to global quantum
correlations~\cite{liu-prl-2007}.  
Mathematical approaches based on symmetry principles and generalized
Pauli constraints characterize the compatibility of reduced
density matrices through constraints on their eigenvalue
spectra~\cite{altunbulak,christandl}, revealing the 
highly structured nature of the quantum marginal problem.
These studies highlight that RDMs are not merely arbitrary local objects, but
encode structured information about the global state.

Once marginal compatibility is established, the central question one should ask
next is whether a global quantum state is uniquely fixed by its marginals,
namely, the reduced density matrices. Pioneering work on three-party pure
states showed that two-body RDMs can uniquely determine the entire
state~\cite{Linden_Wootters,Diosi-2004,sun-2021}.  This insight was extended to
larger systems, where it was shown that, almost all pure states are uniquely
specified by only a subset of their
RDMs~\cite{Linden_PRL,Lyons_PRL,chen-pra-2013,Wyderka_PRA}.  
Remarkably, it was
also shown that genuinely multipartite entanglement can emerge 
even when all reduced density matrices are separable
~\cite{Miklin-pra-2015}.  At the same time, detailed studies of
two-party marginals revealed that specific correlation structures can lead to
fundamentally non-unique global reconstructions, even when all local marginals
are fixed~\cite{chen-pra-2014}. Complementary separability criteria were
developed to identify when multipartite states admit decompositions into
lower-order separable components~\cite{Xu2020,Wyderka2020,zhang-jphysa-2024},
clarifying the structure hidden in reduced data.  These ideas have also been
explored experimentally, with multipartite entangled state reconstruction from
reduced data demonstrated across several quantum
platforms~\cite{dogra-pra-2015-1,
dogra-pra-2015-2,Tao_Xin_PRL_2017,micuda-optica-2019}.

Semidefinite programming(SDP) has emerged as an important computational
framework in quantum information, aiding in quantum state estimation,
entanglement detection, and unearthing measurement
incompatibility~\cite{mironowicz-2024,SDP_review_Tavakoli}.  In the context of
the QMP, systematic hierarchies of SDP tests have been developed to detect
entanglement from local data and to certify compatibility of RDMs with a global
quantum state~\cite{Navascues-quantum-2021,aloy-2021}.  Despite their
advantages, SDP-based approaches are poorly suited to marginal inference, since
they are highly sensitive to noise, treat quantum states as generic elements of
a convex set rather than as structured physical objects, and often cannot
resolve non-uniqueness when several global states share the same reduced
data~\cite{cavalcanti-sdp,Harrow2019}.  Moreover, SDP hierarchies scale
exponentially with system size, making exact marginal certification infeasible
beyond very small systems~\cite{yu-2021}.  These limitations of SDP-based
approaches to the QMP have motivated the search for methods that can exploit
state structure, tolerate noise, and operate directly at the level of reduced
data.

Against this backdrop, machine-learning(ML) approaches have rapidly become a
useful tool in quantum information science, from parameter estimation and
quantum measurement to protocol design and hardware-level
control~\cite{ML-review,alberto-entang-brazil-2025}.  In the context of
entanglement, ML techniques have been used to construct multi-qubit
witnesses~\cite{Vintskevich-ML-2023,entang-ml-pra-2023,zhu-prapplied-2023},
classify multipartite
entanglement~\cite{ml-ent-werner,siamese-prapp-2024,svm-pla}, and certify
nonclassical correlations on noisy intermediate-scale quantum (NISQ)
platforms~\cite{mahdian2025ML,ml-njp-2025,jorawar_2025,Gulati2025,ml-ent-npj-2025,svm-scirep,lin-pra-2026}.
Several studies have demonstrated that neural networks can infer entanglement
properties directly from measurement data, bypassing full state
tomography~\cite{koutny-sciadv-2023,ann-ent-scirep-2024,wang-entropy-2025}.
However, these efforts predominantly treat entanglement characterization as a
forward prediction problem (assuming a well-defined underlying quantum state),
rather than confronting the deeper question of whether reduced data uniquely
specify any global state at all.

Four-qubit pure states exhibit a rich entanglement landscape, comprising
multiple inequivalent SLOCC families and a large number of finer entanglement
classes~\cite{Verstraete,dietrich-2022,D_li,li-pra-2012,giordano-prr-2022}.  We
therefore focus on four qubits as the smallest setting in which
reconstructability, entanglement class, and marginal order already intertwine
in a genuinely nontrivial way.  This makes four-qubit systems a compact testbed
to probe how global quantum structures might be  encoded in reduced data.

In this work, we show that global-state reconstructability is encoded far more
compactly in reduced density matrices than is usually assumed, and that neural
networks can extract this structure directly from marginal data, even in the
presence of realistic noise. This recasts the quantum marginal problem as an
inference task: instead of solving a new constrained optimization problem for
each instance, the network learns the reconstructability structure and the
corresponding inverse maps from data. Our approach departs fundamentally from
most machine-learning studies of entanglement. Rather than estimating
properties of a presumed global state, we address the marginal problem itself:
whether a given set of reduced density matrices uniquely determines a
compatible global quantum state, and whether that state can be explicitly
constructed when uniqueness holds.  By grounding the learning framework in SDP
certification, we ensure that the networks are trained on a physically defined
reconstructability boundary, rather than relying on ad-hoc correlations in the
data.  We benchmark the method on a four-qubit nuclear magnetic resonance
quantum processor, showing that marginal-based neural inference remains
accurate for experimentally prepared entangled states under intrinsic
phase-damping noise.  
%%%%%%%%%%%%%%%%%%%%%%%%%%%%%%%%%%%%%%%%%%
\section{Results} 
\begin{figure*}
\includegraphics[scale=1]{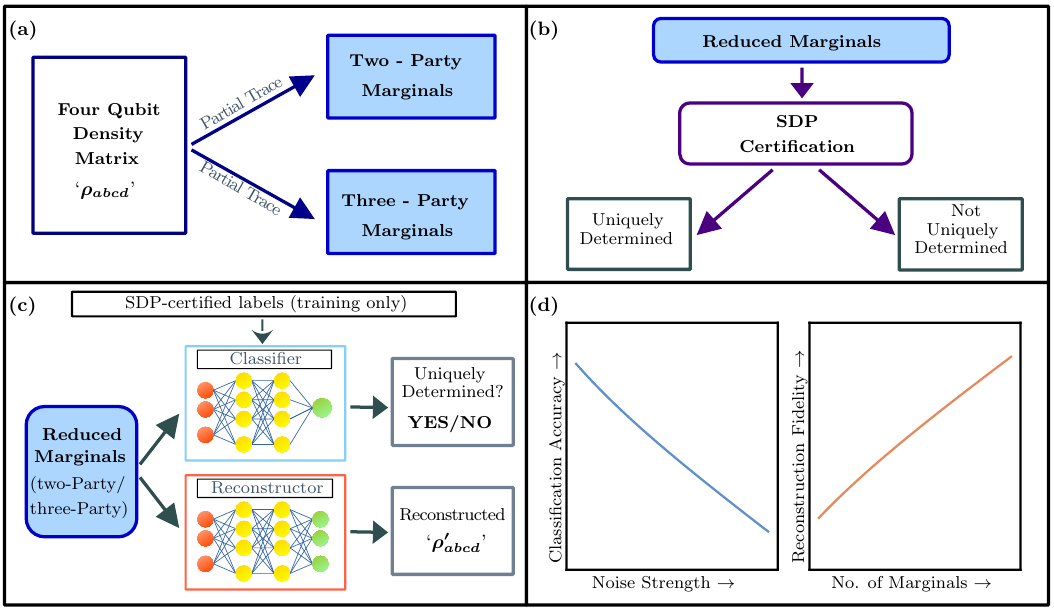} \caption{ANN framework
for the four-qubit quantum marginal problem.  (a) Global four-qubit
states are accessed only through their two-body and three-body reduced
density matrices.  (b) Reconstructability analysis using semidefinite
programming (SDP), which determines whether a given set of marginals
uniquely specifies a compatible global state.  (c) Neural-network
framework comprising two tasks: classification of reconstructability
directly from marginals, and constructive reconstruction of four-qubit
density matrices when uniqueness holds.  (d) ANN performance analysis
using classification accuracy and state-reconstruction fidelity as
functions of noise strength and available marginal information.}
\label{figure1} 
\end{figure*}

We first establish the reconstructability landscape of four-qubit states
using semidefinite programming and then show how neural networks exploit this
structure to infer and reconstruct global quantum states from reduced density
matrices.
\subsection{ANN model for the quantum marginal problem}
\label{ann-model}
The central question of this work is whether reduced density
matrices uniquely determine a compatible global quantum
state, and if so, whether that state can be explicitly
recovered from reduced data alone. We address this question
in four-qubit systems, which already display nontrivial
marginal structure while remaining accessible to systematic
numerical analysis.
	
As illustrated in Fig.~\ref{figure1}, we begin from global four-qubit states
and construct two-party and three-party reduced density matrices, which serve
as the only inputs to our ANN models. From this reduced data, we define two
operational tasks.  The first is a decision task to determine whether a given
set of marginals uniquely specifies a compatible global state and the second is
a constructive task to recover a four-qubit density matrix consistent with the
marginals.

Reconstructability is certified using SDP, which determines whether a given set
of marginals uniquely determine a compatible global quantum state.  These SDP
results define the reconstructability landscape and provide the reference
labels used throughout this work. Neural networks are then trained to learn
this structure, using only reduced density matrices. They learn to classify
whether marginals are uniquely compatible, and to reconstruct the corresponding
global state when they are.  This framework links certified marginal uniqueness
to explicit state inference from reduced data.

%%%%%%%%%%%%%%%%%%%%%%%%%%%%%%%%%%%%%%%%%%%%%%%%%%%%%%%%%%%%%%%%

\begin{table}[h]
\centering
			\renewcommand{\arraystretch}{1.4}
			\setlength{\tabcolsep}{8pt}
			\begin{tabular}{l|c|c}
				\hline
				\hline
				SLOCC Class & 3-Party Marginals & 2-Party Marginals \\
				\hline
				Separable      & Yes & Yes \\
				Biseparable    & Yes & Yes \\
				Triseparable   & Yes & Yes \\
				$G_{abcd}$     & Yes & No  \\
				$L_{abc_2}$    & Yes & Yes \\
				$L_{a_2b_2}$   & Yes & Yes \\
				$L_{ab_3}$     & Yes & Yes \\
				$L_{a_4}$      & Yes & Yes \\
				$L_{a_20_{3\oplus \bar{1}}}$ & Yes & Yes \\
				$L_{0_{5\oplus\bar{3}}}$    & Yes & Yes \\
				$L_{0_{7 \oplus\bar{1}}}$    & Yes & No  \\
				$L_{0_{3\oplus\bar{1}}0_{3\oplus\bar{1}}}$ & Yes & No \\
				\hline
			\end{tabular}
\caption{SDP-certified reconstructability of four-qubit states from reduced
marginals summarized across different SLOCC entanglement families. The
table lists whether each class is uniquely reconstructible from
two-party and three-party marginals, with `Yes' (`No') indicating
reconstructible (non-reconstructible) cases. Among these families, two
exceptional classes arise: the $G_{abcd}$ (A1.1) class, which is
non-reconstructible from both two-party and three-party marginals, and
the $L_{a_4}$ (La1) class, which is non-reconstructible from two-party
marginals but becomes reconstructible when three-party marginals are
used.
}
\label{tab:slooc_reconstructability}
\end{table}
	
\begin{figure}
\centering
\includegraphics[scale=1]{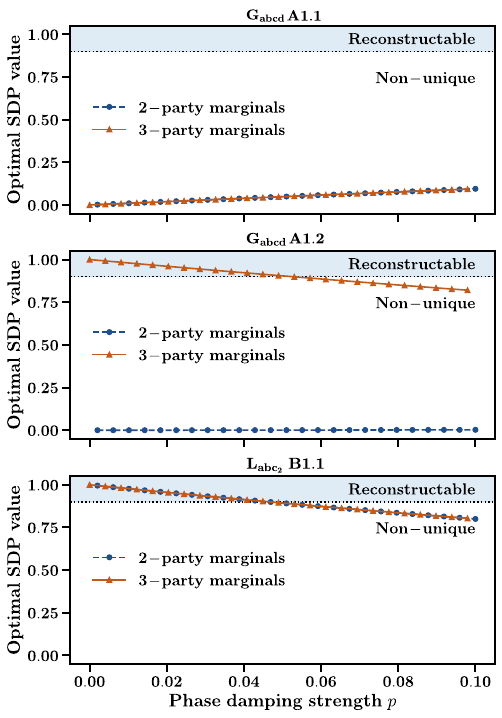}
\caption{SDP results for reconstructability of four-qubit states from
reduced marginals. The effect of phase-damping noise on
reconstructability is shown for three representative classes of four qubits,
quantified by the optimal SDP value as a function of the noise strength
$p$.} 
\label{figure2}
\end{figure}

The resulting landscape, shown in Table~\ref{tab:slooc_reconstructability}, is
highly structured.  Reconstructability depends strongly on
both entanglement class and marginal order. Three-party
marginals uniquely fix a wide range of genuinely
multipartite entangled states, whereas two-party marginals
often fail to determine a unique global state.  Distinct
patterns emerge across SLOCC families and exceptional
subclasses: some are robustly reconstructible, while others
are intrinsically non-unique even in the absence of noise.
Separable and partially separable states obey markedly
different marginal constraints from genuinely entangled
families, underscoring the role of higher-order
correlations.  The effect of noise is shown in
Fig.~\ref{figure2}. Under phase damping, 
he optimal SDP value
decreases smoothly, revealing regimes in which
uniqueness persists and others in which reconstructability
collapses. States whose global structure is encoded in
higher-order marginals remain reconstructible under
decoherence, while states dominated by pairwise information
are significantly more fragile.
	%%%%%%%%%%%%%%%%%%%%%%%%%%%%%%%%%%%%%%%%%%%%%%%%%%%%%%%%%%%%%%%%%%%%%%%
\subsection{Neural classification of marginal
reconstructability}

\begin{figure}[ht]
\includegraphics[scale=1]{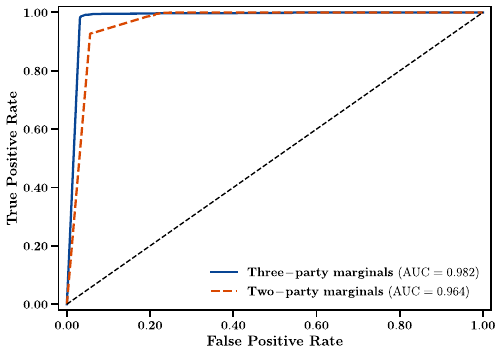} 
\caption{Receiver operating characteristic (ROC) curves for classifiers
based on three-party marginals (blue solid curve) and two-party
marginals (orange dashed curve), evaluated at fixed phase damping noise
strength $p = 0.03$. The curves show the true positive rate as a
function of the false positive rate, with the diagonal dashed line
indicating random guessing. The area under the curve (AUC) is $0.982$
for three-party marginals and $0.964$ for two-party marginals.} 
\label{figure3} 
\end{figure}

We train neural classifiers to decide marginal reconstructability using
the reduced density matrices as input, with no access to the
full four-qubit state.  Two models are studied: one receiving three-party
marginals and one receiving only two-party marginals. In both cases, the target
labels are assigned by SDP certification.

Figure~\ref{figure3} summarizes the performance of the marginal-based
classifiers through receiver operating characteristic (ROC) analysis. Both
classifiers distinguish reconstructible from non-reconstructible states with
high fidelity, yielding ROC curves that lie well above the random-guessing
baseline. The classifier based on three-party marginals achieves superior
discrimination, consistent with the greater information content available in
higher-order reduced density matrices. These trends are reflected
quantitatively in the classification accuracies reported below.

Consistent with the ROC analysis, the classifier based on three-party
marginals achieves accuracies of $99.93\%$, $99.93\%$, and $98.40\%$ on the
training, validation, and test datasets, demonstrating excellent
generalization.  The corresponding classifier based on two-party marginals
achieves attains accuracies of $97.05\%$, $96.97\%$, and $95.36\%$, indicating
consistently lower, but still robust, predictive performance.

The ROC curves in Fig.~\ref{figure3} further quantify the discriminative
ability of the two models.  The classifier based on three-party marginals
achieves an area under the curve (AUC) of 0.982, compared with 0.964 for the
two-party classifier. The superior AUC obtained with three-party marginals
reflects the greater information content carried by higher-order reduced
density matrices, making them more effective for 
determining reconstructability.

Classifiers based on the full four-qubit global state were also developed and
evaluated. Details of their architecture, training, and performance are
given in the Supplementary Material.

%%%%%%%%%%%%%%%%%%%%%%%%%%%%%%%%%%%%%%%%%%%%%%%%%%%%%%%%%%%%%%%%%%%%%%%
\subsection{Neural reconstruction of four-qubit states from reduced marginals}
\begin{figure}
\centering
\includegraphics[scale=1]{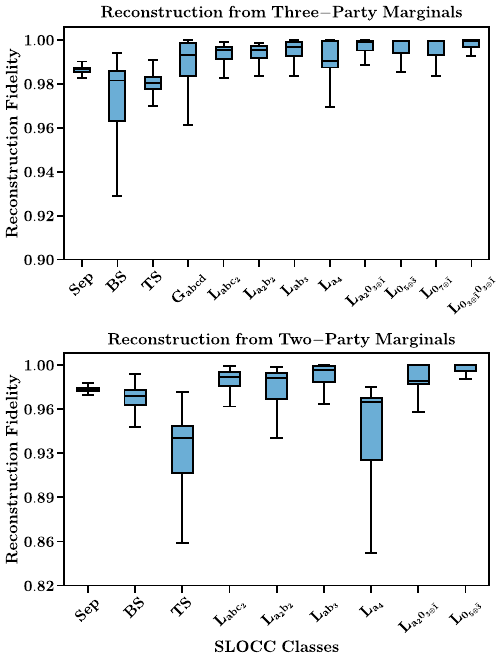}
\caption{Neural reconstruction of four-qubit states from reduced density
matrices. Upper panels show reconstruction using all three-party
marginals, while lower panels show reconstruction using all two-party
marginals. The plots display class-resolved fidelity distributions
across SLOCC families for SDP-certified reconstructible states, with
each box plot compiled from $10^4$ samples per class including both
pure and noisy states. All reconstructions used reduced density
matrices alone, without access to the global four-qubit state.} 
\label{figure4}
\end{figure}

We now turn from deciding reconstructability to explicitly recovering global
four-qubit states from RDMs alone. Reconstruction networks were trained only on
SDP-certified reconstructible states, so that the learning problem corresponds
to a well-defined inverse map from marginals to global states. Two settings
were studied: reconstruction from complete sets of three-party marginals and
from complete sets of two-party marginals.
	
Figure~\ref{figure4} compares neural reconstruction fidelity across
SDP-certified reconstructible SLOCC classes using complete sets of two-party
and three-party reduced density matrices. Reconstruction from three-party
marginals is uniformly reliable, with fidelity distributions concentrated close
to unity for all entanglement classes. In contrast, reconstruction from
two-party marginals is markedly more class dependent. Although several classes
remain accurately reconstructed, others exhibit broader fidelity distributions
and lower median fidelities, reflecting the reduced information content of
pairwise correlations. The systematic improvement obtained with three-party
marginals demonstrates that higher-order reduced correlations capture the
global structure of four-qubit states far more effectively than two-party
marginals alone.

%%%%%%%%%%%%%%%%%%%%%%%%%%%%%%%%%%%%%%%%%%%%%%%%%%%%%%%%%%%%%%
\subsection{Reconstruction from incomplete marginal sets}
\begin{figure*}
\centering
\includegraphics[scale=1]{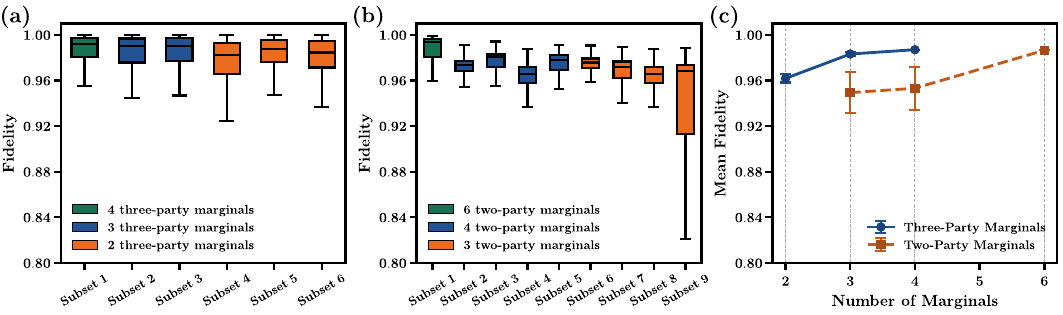}
\caption{Reconstruction fidelity as a function of marginal information.
(a) Fidelity distributions for neural reconstruction of four-qubit
states from subsets of three-party marginals, grouped by the number of
marginals used. Each box plot corresponds to a distinct marginal
configuration.  (b) Corresponding fidelity distributions for
reconstruction from subsets of two-party marginals.  (c) Mean
reconstruction fidelity versus the number of available marginals,
averaged across all marginal configurations within each group. Error
bars indicate the standard deviation across configurations.  All
fidelities are computed on sets of $10^4$ states drawn from all SLOCC
classes used in training, including both pure and noisy states.  The
results reveal clear information thresholds in marginal-based
reconstruction.}
\label{figure5} 
\end{figure*}

We next ask how much reduced information is actually needed to reconstruct a
	four-qubit state. Rather than supplying complete marginal sets, we
	systematically restrict the inputs and train reconstruction networks on
	selected subsets of two-party and three-party reduced density matrices. This
	isolates which marginals are redundant, which are sufficient, and where
	reconstruction fails.
	
	Fig.~\ref{figure5} shows a clear separation between three-body and two-body
	information.  For subsets of three-party marginals (Fig.~\ref{figure5}(a)),
	high fidelities persist across many configurations, even when only a few
	marginals are provided, indicating strong redundancy among three-body
	reductions. In contrast, reconstruction from two-party subsets
	(Fig.~\ref{figure5}(b)) deteriorates rapidly as marginals are removed, with
	broader fidelity distributions and reduced medians, reflecting the limited
	ability of pairwise correlations to pin down global quantum structure.  The
	averaged trends in Fig.~\ref{figure5}(c) reveal a monotonic scaling with
	marginal number: fidelity improves steadily as more marginals are added, with
	three-party subsets consistently outperforming two-party subsets at fixed
	count. These results identify operational information thresholds in the
	marginal problem. They show that global four-qubit structure is often encoded
	compactly in small sets of higher-order marginals, while also delineating the
	sharp limits of inference from incomplete reduced data.
	%%%%%%%%%%%%%%%%%%%%%%%%%%%%%%%%%%%%%%%%%%%%%%%%%%%%%%%%%%%%%%%%%
	
In addition to the reconstruction models based solely on two-party or
three-party marginals, a mixed marginals approach was also investigated. In
this case, the input consisted of one three-party marginal, $\rho_{abc}$,
together with three two-party marginals, $\rho_{ab}$, $\rho_{bd}$, and
$\rho_{cd}$. The objective was to reconstruct the full four-qubit density
matrix using information from marginals of different orders.
		
\begin{figure*}[t]
\centering
\includegraphics[scale=1]{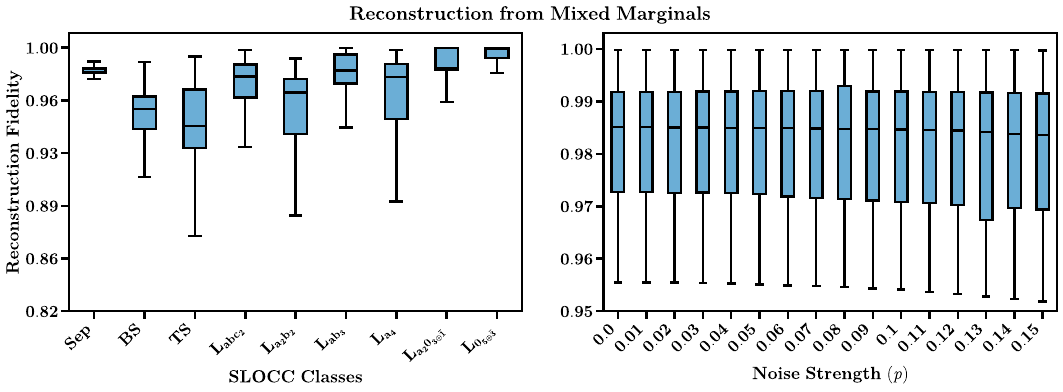}
\caption{ANN analysis of four-qubit states using mixed marginals. Left
panels show state-wise performance across different SLOCC classes, with
each distribution compiled from $10^4$ test samples per dataset. Right
panels show performance as a function of phase-damping strength $p$,
evaluated on independent test sets of $10^4$ states at each noise
value.
}
\label{mixed_marginals_res}
\end{figure*}

The model demonstrated excellent reconstruction performance, with average
fidelities exceeding $99\%$ on both the training and validation datasets. The
corresponding results are presented in Fig.~\ref{mixed_marginals_res}.

\subsection{Representative reconstructions and intrinsic failure modes}
\begin{figure*} 
\centering 
\includegraphics[scale=1]{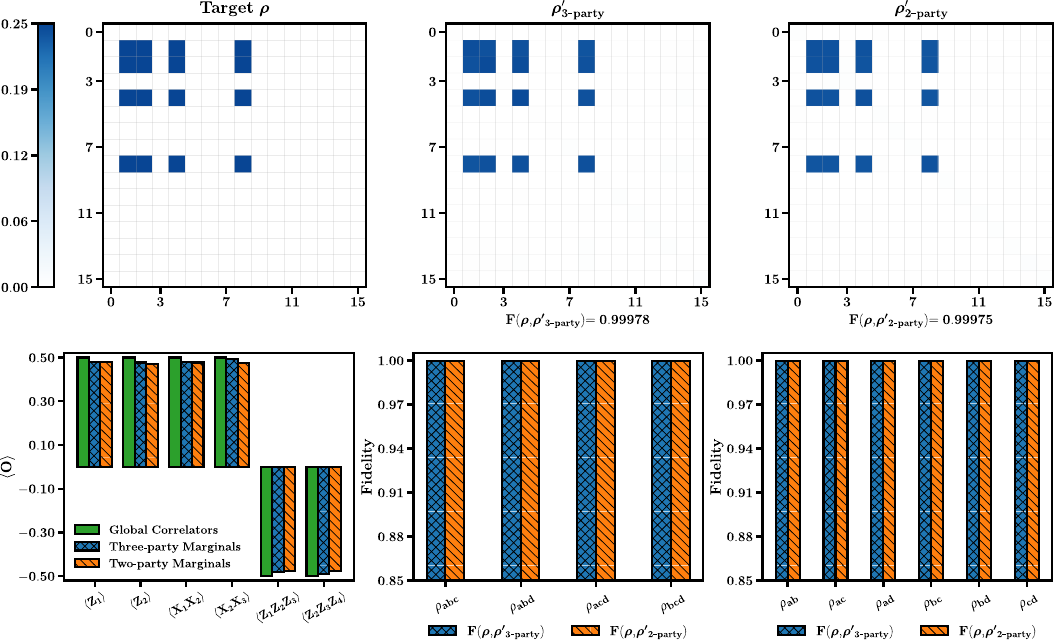}
\caption{Reconstruction analysis for a representative four-qubit
W state.  (a) Density matrix of the target state, $\rho$, and the
reconstructed density matrices $\rho'_{\mathrm{3\text{-}party}}$ and
$\rho'_{\mathrm{2\text{-}party}}$, obtained using all available
three-party and two-party marginals, respectively. A single color bar
(shown on the left) is used for all three density matrices.  (b)
Preservation of physical properties under reconstruction. The left
panel shows selected Pauli correlators evaluated for the global target
state $\rho$, the state reconstructed from three-party marginals
$\rho'_{\mathrm{3\text{-}party}}$, and the state reconstructed from
two-party marginals $\rho'_{\mathrm{2\text{-}party}}$. The middle and
right panels show the fidelities between the reduced marginals of the
reconstructed states $\rho'_{\mathrm{3\text{-}party}}$ and
$\rho'_{\mathrm{2\text{-}party}}$ and the corresponding exact reduced
marginals, demonstrating high reconstruction accuracy.} 
\label{figure6} 
\end{figure*}

To illustrate the reconstruction at the level of individual states,
	Fig.~\ref{figure6} shows a state-resolved reconstruction for a representative
	four-qubit W state. Panel (a) shows the reconstructed density matrices obtained
	from complete sets of three-party and two-party marginals, together with their
	deviations from the target state. In both cases the residuals are small and
	largely structureless, indicating faithful recovery of the global state.  Panel
	(b) evaluates the reconstructions at the level of physical observables.
	Selected Pauli correlators and reduced marginals computed from the
	reconstructed states closely match those of the target, confirming that the
	networks preserve both global coherence and local marginal constraints. High
	fidelity here therefore corresponds to physically meaningful reconstruction,
	not merely a high-overlap fit.
	
	For intrinsically non-reconstructible families, the networks converge to
	different marginal-consistent but globally inequivalent states.  The
	corresponding low fidelities directly reflect the underlying non-uniqueness of
	the marginal problem.  A detailed analysis of GHZ-type failure modes,
	illustrating the systematic loss of global coherence despite fixed marginals,
	is presented in the Supplementary Material.
	\subsection{Experimental validation on a four-qubit NMR processor}
	%%%%%%%%%%%%%%%%%%%%%%%%%%%%%%%%%%%%%%%%%%%
\begin{figure}[ht] 
\centering 
\includegraphics[scale=1]{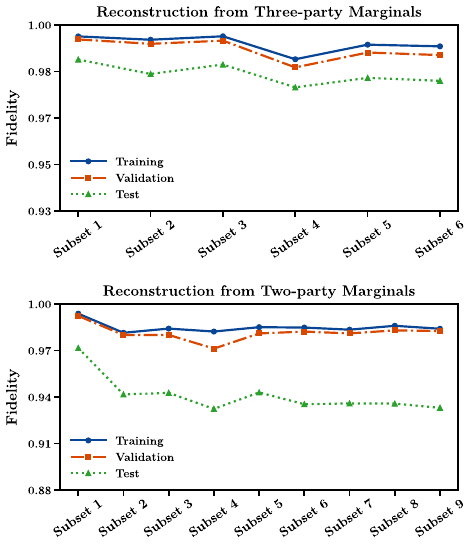}
\caption{Average reconstruction fidelity on the training, validation, and
independent test sets for models trained on different subsets of
reduced density matrices. The upper panel shows reconstruction from
three-party marginals (six marginal subsets), and the lower panel shows
reconstruction from two-party marginals (nine marginal subsets). Test
fidelities remain close to training and validation performance,
demonstrating stable learning and strong generalization. Fidelities are
evaluated on independent test sets of $10^4$ states drawn from all
SLOCC classes used in training.  }
\label{figure7}
\end{figure}

%%%%%%%%%%%%%%%%%%%%%%%%%%%%%%%%%%%%%%%%%%%
%%%%%%%%%%%%%%%%%%%%%%%%%%%%%%%%%%%%%%%%%%%
\begin{figure*} 
\centering 
\includegraphics[scale=1]{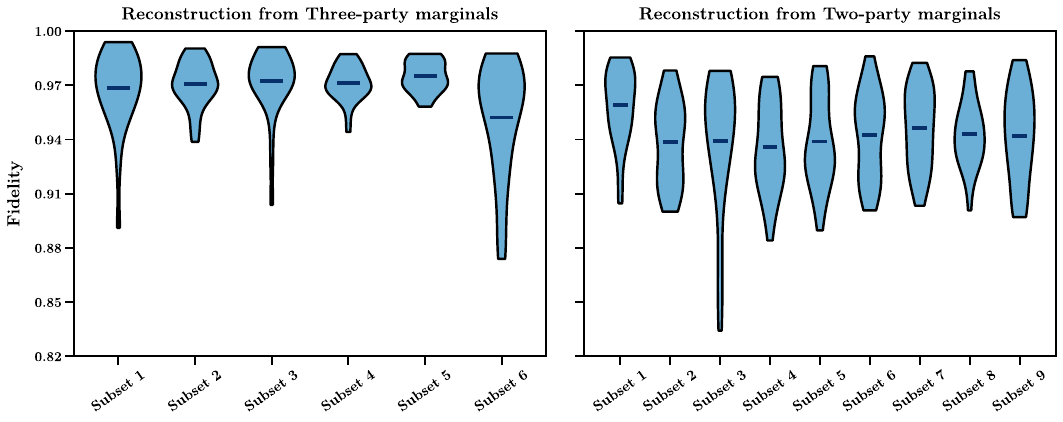}
\caption{Reconstruction fidelities for 30 experimentally prepared
four-qubit states shown as violin plots. Each violin represents how the
fidelity values are distributed for a given marginal subset. The left
group corresponds to three-party marginals, while the right group
corresponds to two-party marginals. Wider regions indicate where values
are more concentrated, and the central marker shows the mean fidelity.}
\label{figure8}
\end{figure*}

%%%%%%%%%%%%%%%%%%%%%%%%%%%%%%%%%%%%%%%%%%%
	We tested the marginal-based neural framework on a four-qubit nuclear magnetic
	resonance (NMR) quantum processor, where the qubits are encoded in the
	${}^{13}$C nuclear spins of labeled trans-crotonic acid. Intrinsic transverse
	T$_2$ NMR relaxation provides a natural source of phase-damping noise, making
	this platform well suited for probing the robustness of marginal-based
	inference under realistic experimental conditions.
	
	Two experimental datasets were prepared. The first comprised seven
	representative four-qubit states spanning both reconstructible and
	non-reconstructible classes, each implemented multiple times with varying
	fidelities to probe robustness. The second consisted of 30 randomly generated
	entangled states, all certified by SDP to be reconstructible from their two-
	and three-party marginals. In each case, full four-qubit quantum state
	tomography was performed, and the corresponding reduced density matrices were
	supplied to the trained networks.
	
The classifiers were tested on the first experimental dataset and
correctly identified the reconstructability class of every state, achieving a
classification accuracy of $100\%$. The detailed classification results,
together with the corresponding experimental state fidelities, are provided in
the Supplementary Material.

Fig.~\ref{figure7} compares reconstruction fidelities with those obtained on
training, validation, and independent numerical test sets.  Experimental
performance lies within the numerical envelope for both three-party- and
two-party-based models, showing that the learned inverse maps remain stable in
the presence of decoherence and control imperfections.  Figure~\ref{figure8}
reports blind reconstructions of the 30 random experimental states. High
fidelities are obtained consistently across states and marginal configurations,
demonstrating that global four-qubit states can be faithfully recovered in an
experiment, from reduced marginals alone. Additional experimental data, state
constructions, and pulse-sequence details are provided in the Supplementary
Material.
	
\subsection{Noise Analysis of Neural Network Models}
Since practical quantum-state reconstruction inevitably involves noisy
measurements, we finally examine the robustness of the learned
marginal-inference models under realistic decoherence.
To evaluate robustness under realistic experimental conditions, we trained
marginal-based classifiers and reconstructors on datasets comprising pure
states, phase-damped, amplitude-damped, depolarised states, and a combined
dataset containing all noise types. All classifier models achieved high
training accuracies, exceeding $98\%$ for three-party marginals and $95\%$ for
two-party marginals.

\begin{figure}[tbh!] \centering
\includegraphics[scale=1]{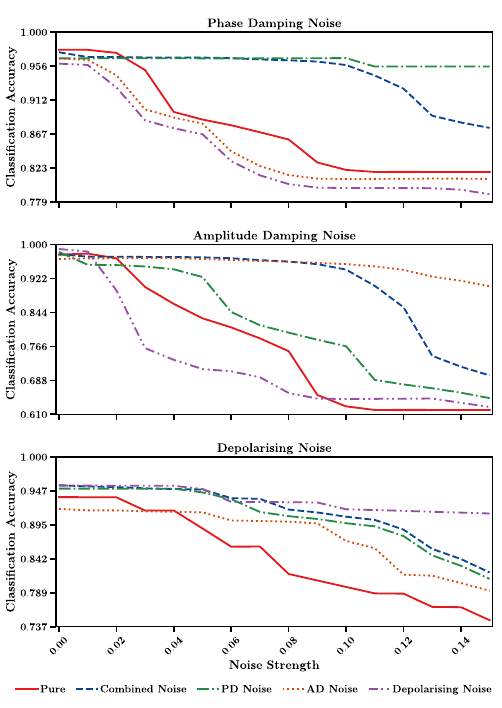}
\caption{
Classification accuracy of neural-network classifiers predicting the
reconstructability of four-qubit states from three-party marginals
under phase-damping (top), amplitude-damping (middle), and depolarising
(bottom) noise. Curves correspond to models trained on pure,
noise-specific, and combined-noise datasets, evaluated on independent
test sets of $10^4$ states at each noise level. Training was performed
up to $p=0.05$; higher noise strengths assess generalization beyond the
training regime. Performance is highest when the training and test
noise distributions are matched.}
\label{noisevsacc} 
\end{figure}

Figure~\ref{noisevsacc} summarizes the classification performance of models
trained under different noise conditions and evaluated on independent test sets
of $10^4$ states. Models trained solely on pure states perform well at low
noise levels but degrade rapidly as the noise strength increases. In contrast,
models trained on noisy or combined-noise datasets generalize substantially
better, with the highest accuracies obtained when the training and test noise
distributions match. Combined-noise training provides the greatest robustness
across different noise models, whereas depolarising-noise models exhibit weaker
performance owing to both the limited training range ($p\leq0.02$) and the more
disruptive nature of depolarising noise. Similar trends are observed for
classifiers based on two-party marginals (see Supplementary Material).

For state reconstruction, separate networks were trained using all available
two-party or three-party marginals to learn the mapping from noisy reduced
density matrices to the corresponding clean four-qubit density matrix. All
reconstruction models achieved fidelities exceeding $98\%$ on the training and
validation datasets.

\begin{figure}[t] \centering
\includegraphics[scale=1]{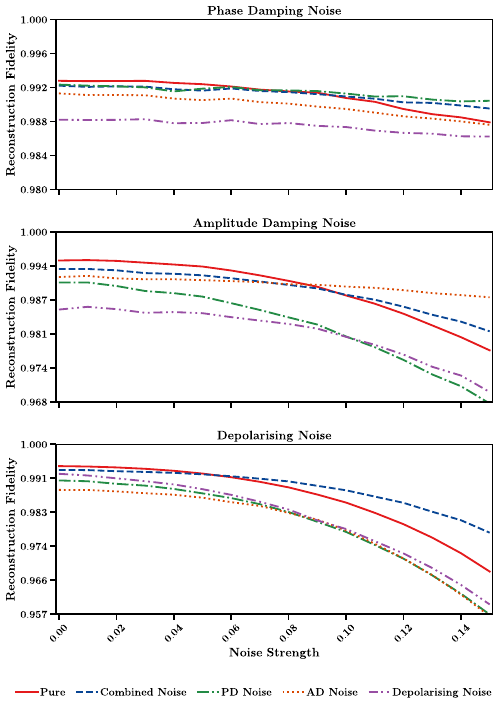}
\caption{
Reconstruction fidelity of four-qubit states from three-party
marginals under phase-damping (top), amplitude-damping (middle), and
depolarising (bottom) noise. Curves correspond to models trained on
pure, noise-specific, and combined-noise datasets, evaluated on
independent test sets of $10^4$ states at each noise level. Training
was performed up to $p=0.05$; higher noise strengths probe
generalization beyond the training regime. Models trained on noise
matching the test distribution achieve the highest reconstruction
fidelity.
} 
\label{noisevsfid} 
\end{figure}

The reconstruction fidelities on independent test sets are shown in
Fig.~\ref{noisevsfid}. As in the classification task, models trained on
noise-matched datasets consistently outperform those trained solely on pure
states at moderate and high noise levels, while combined-noise training
provides robust performance across different noise channels. Compared with
classification, however, reconstruction fidelity degrades more gradually with
increasing noise, indicating superior generalization. This difference reflects
the nature of the two tasks: classification requires learning a decision
boundary that shifts under noise, whereas reconstruction is a regression
problem that learns a continuous mapping and is therefore inherently more
robust to perturbations. Even on an experimental dataset of 30 four-qubit
states, the reconstruction models maintained fidelities exceeding $94\%$.
State-resolved experimental fidelities are provided in the Supplementary
Material.
	
%%%%%%%%%%%%%%%%%%%%%%%%%%%%%%%%%%%%%%%%%%%%%%%%%%%%%%%%%%%%%%
\section{Discussion} 
Most machine-learning studies in quantum information aim
to estimate properties of an assumed global state from partial data. We instead
confront the quantum marginal problem itself i.e., when does local quantum data
suffice to determine a global state, 
and what governs that sufficiency?

By combining SDP certification with neural inference, we show that
marginal reconstructability is not generic, but is systematically organized by
the interplay between the available reduced density matrices and the underlying
multipartite entanglement structure.

In contrast to three-qubit systems, four-qubit states support a much
richer entanglement landscape, with distinct SLOCC families and subclasses
exhibiting qualitatively different reconstructability properties, ranging from
robust uniqueness to intrinsic non-uniqueness.
The strong contrast between two-
and three-body marginals reveals a clear informational hierarchy, of
higher-order  marginals retaining collective constraints associated with
genuinely multipartite correlations, and pairwise marginals frequently
underdetermining the global quantum state.  Reconstructability is thus a
meaningful probe of the structure of multipartite entanglement.  

Neural
networks trained solely on reduced density matrices learn to infer whether
uniqueness holds and, when it does, to reconstruct the global state itself. The
close correspondence between SDP structure and classifier performance indicates
that the networks are capturing compatibility constraints intrinsic to the
marginal problem, not superficial statistical features.
The presence of  noise
reshapes this structure without erasing it.  Specifically, phase damping
drives a smooth degradation of both uniqueness and reconstruction fidelity,
distinguishing regimes where reconstructability is lost from
those where it remains but becomes harder to infer. The ability of the models
to generalize across entanglement families and beyond the training noise window
suggests that they learn structural features of noisy marginal data rather than
a narrow noise model.  State-resolved reconstructions further confirm that high
fidelity reflects preservation of genuine quantum correlations and local
compatibility, while systematic failures for GHZ-type families can be traced
back to intrinsic marginal non-uniqueness.
A detailed comparison of the computational performance of the SDP-based
approach and the proposed neural network models, including inference times for
individual instances, is provided in the Supplementary Material.

Restricting the available marginals reframes the quantum marginal problem in
terms of informational economy. Several subsets of three-party marginals
support faithful reconstruction, revealing substantial redundancy in how global
four-qubit structure is distributed across higher-order marginals. In contrast,
restricted two-party sets degrade rapidly, implying their propensity to
underdetermine the global state.  The neural networks are able to identify
operational thresholds separating sufficient, redundant, and insufficient
marginal configurations.  Testing the networks on experimental NMR data
validate that networks trained on numerical data can accurately classify
reconstructability and reconstruct global states directly from experimentally
measured marginals despite decoherence and control imperfections.  Our results
position marginal-based learning as a practical route to compressed quantum
tomography, and as a framework for exploring how global quantum structure is
encoded, redundantly but unevenly, in local data.  
\section{Methods}
\label{meth} 
\subsection{Four-qubit state generation and noise models} 
Training and test data were generated from the standard SLOCC classification of
four-qubit pure states introduced by Verstraete \textit{et
al.}~\cite{Verstraete}, which organizes genuinely entangled states into
nine inequivalent families. 

Stochastic local operations and classical communication (SLOCC) provide the
natural equivalence relation for multipartite entanglement. Two pure states
belong to the same SLOCC class if they can be converted into one another with
nonzero probability using invertible local operations assisted by classical
communication. States within a class therefore possess the same qualitative
type of multipartite entanglement, even though their amplitudes and local bases
may differ.

Unlike the three-qubit case, which contains only a finite number of
inequivalent entanglement classes, four-qubit systems exhibit infinitely many
SLOCC classes. Verstraete \textit{et al.} showed that these can nevertheless be
organized into nine canonical entanglement families, denoted by:
$G_{abcd}$,
$L_{abc_2}$,
$L_{a_2b_2}$,
$L_{ab_3}$,
$L_{a_4}$,
$L_{a_20_{3\oplus\bar1}}$,
$L_{0_{5\oplus\bar3}}$,
$L_{0_{7\oplus\bar1}}$, and
$L_{0_{3\oplus\bar1}0_{3\oplus\bar1}}$.

Each family is represented by a canonical state depending on a small number of
complex parameters. Different parameter values generally correspond to
inequivalent SLOCC classes, while special algebraic relations among the
parameters produce singular subclasses possessing distinct entanglement
structure. Consequently, each family contains several inequivalent true SLOCC
classes, yielding 49 inequivalent classes in total after refinement.

For our dataset generation, states were sampled directly from these 49 refined
SLOCC classes by assigning random complex values to the free family parameters
while enforcing the algebraic constraints defining each class. Explicit
canonical forms and the parameter constraints used to generate every class are
given in the Supplementary Material.

Subsequent work refined these nine canonical families 
into 49 inequivalent true SLOCC classes~\cite{D_li}, 
which formed the basis of our dataset generation.
In addition, we
included fully separable, biseparable, and triseparable states to ensure broad
coverage of the four-qubit entanglement landscape.

States from each SLOCC class were constructed by assigning random complex
values to the defining family parameters, subject to the algebraic constraints
of the class, and normalized. Separable states were generated as tensor
products of four independently sampled single-qubit pure states drawn uniformly
from the Bloch sphere. Biseparable and triseparable states were generated by
enforcing separability across the appropriate bipartitions, with the remaining
subsystems entangled.  To remove
basis bias and enforce local-unitary invariance, independently drawn random
unitaries from $SU(2)$ were applied to each qubit of every generated state.
This produces physically equivalent but locally distinct realizations within
each entanglement class.
	
To model realistic experimental imperfections, all states were subjected to
single-qubit phase-damping noise channels, \begin{eqnarray} \mathcal{E}_p(\rho)
	&=& \sum_{i=0,1} K_i \rho K_i^\dagger \nonumber \\ K_0 &=&
	\begin{bmatrix} 1 & 0 \\ 0 & \sqrt{1-p} \end{bmatrix} \quad K_1 =
		\begin{bmatrix} 0 & 0 \\ 0 & \sqrt{p} \end{bmatrix}
\end{eqnarray} applied independently to each qubit. The damping parameter $p$
was varied over a fixed range to generate mixed-state ensembles reflecting
decoherence processes relevant to NMR platforms.
	
To evaluate robustness under different noise conditions, neural networks
were trained on datasets comprising pure states, pure states mixed with phase-,
amplitude-, or depolarising-noise, and a combined dataset containing all noise
types. For noisy states, the SDP reconstructability threshold was set to 0.9.
Under this criterion, reconstructability remained essentially unchanged up to
noise strengths of $p=0.05$ for phase- and amplitude-damping channels, whereas
for depolarising noise the threshold was satisfied only up to $p\approx0.02$.
Accordingly, training datasets were generated using noise strengths up to $5\%$
for phase and amplitude damping, and up to $2\%$ for depolarising noise.
The Kraus operators
for amplitude damping noise are given by \[ K_0 = \begin{pmatrix} 1 & 0 \\ 0 &
	\sqrt{1-\gamma} \end{pmatrix}, \quad K_1 = \begin{pmatrix} 0 &
\sqrt{\gamma} \\ 0 & 0 \end{pmatrix}.  \]
		
For depolarising noise, the noisy states were generated by applying the
identity operator with probability $(1 - p)$, and the Pauli operators
$\sigma_x$, $\sigma_y$, and $\sigma_z$ each with probability $p/3$, resulting
in an effective randomization of the quantum state. These datasets form the
basis for both SDP labeling and neural-network training.  Details of dataset
sizes, train–validation splits, and task-specific sampling are given in the
Supplementary Material.

%%%%%%%%%%%%%%%%%%%%%%%%%%%%%%%%%%%%%%%%%%%%%%%%%%%%%%%%%
\subsection{SDP formulation of reconstructability} To label states according to
marginal reconstructability, we use semidefinite programming to test whether a
given four-qubit state is uniquely specified a chosen set of reduced density
matrices.  For each target state $\rho = \vert \psi \rangle \langle \psi \vert$
(pure or noisy), we introduce a candidate global density matrix $X$, a $16
\times 16$ Hermitian positive semidefinite matrix with unit trace.
	
Reconstructability is assessed by solving the SDP $$ {\rm minimize} \quad
\mathrm{Tr}(\rho X) $$ subject to \[ X \succeq 0, \quad \mathrm{Tr}(X) = 1,
\quad \mathrm{Tr}_{S \setminus S_j}(X) = \rho_{S_j}, \] where $\rho_{S_j}$ are
the specified two- or three-qubit marginals.  If the SDP returned
$\mathrm{Tr}(\rho X)\approx 1$, indicating that no other global state
compatible with the marginals exists, the state was labeled uniquely
reconstructible.  Two labeling pipelines were implemented: one using all
three-party marginals and one using only two-party marginals.  The SDPs were
implemented in Python using \texttt{CVXPY}  with the \texttt{SCS} solver. To
account for numerical tolerance and experimental noise, states with optimal
value greater than $0.9$ were labeled as reconstructible, and those below as
non-reconstructible. 

For each set of reduced density matrices, reconstructability was certified
using a semidefinite program that searched for a global density matrix
consistent with the prescribed marginals while satisfying the physical
constraints of positivity and unit trace. Uniqueness of the feasible solution
was then used to determine whether the state was reconstructible from the given
marginals. Details of the SDP formulation and results of SDP
reconstructability of all SLOCC classes from RDMs are given in the
Supplementary Material.

%%%%%%%%%%%%%%%%%%%%%%%%%%%%%%%%%%%%%%%%%%%%%%%%%%%%%%%%%%%%
\subsection{Neural network models, inputs, and training} We consider four
learning tasks: classification of reconstructability from three-party
marginals, classification from two-party marginals, reconstruction of
four-qubit states from three-party marginals, and reconstruction from two-party
marginals.
	
For each four-qubit density matrix $\rho_{abcd}$, all two-body marginals
$\{\rho_{ab},\rho_{ac},\rho_{ad},\rho_{bc},\rho_{bd},\rho_{cd}\}$ and
three-body marginals $\{\rho_{abc},\rho_{abd},\rho_{acd},\rho_{bcd}\}$ were
obtained by partial trace. Reduced density matrices were encoded by separating
real and imaginary parts and vectorizing only independent elements, exploiting
Hermiticity. The full four-qubit density matrix was encoded in the same form
for use as the reconstruction target. All marginal-based models therefore
operate exclusively on reduced information, without access to the global state.
	
All models were implemented as feed-forward neural networks using
\texttt{Keras}/\texttt{TensorFlow}. Classification networks were trained to map
reduced marginals directly to a binary reconstructability label. To suppress
redundant inputs and improve training efficiency, univariate ANOVA feature
selection was applied to marginal feature vectors, and the highest-ranked
features were retained. Hidden layers used ReLU activations and a sigmoid
output layer, with binary cross-entropy loss optimized using Adam. Model depth
and width were tuned empirically to balance accuracy and generalization.
	
Reconstruction networks were trained only on SDP-certified reconstructible
states, with reduced marginals as input and the full four-qubit density matrix
as output. Hidden layers used ReLU activations and a linear output layer, with
mean-squared error as the training loss. Since raw network outputs are not
guaranteed to represent physical states, predictions were projected onto the
space of positive semidefinite, unit-trace matrices via $\rho_{\mathrm{pred}}=M
M^\dagger/\mathrm{Tr}(M M^\dagger)$.  Reconstruction performance was quantified
using quantum-state fidelity.
	
All datasets were randomly partitioned into training and validation sets, with
independent test sets used for final benchmarking. Early stopping based on
validation loss was applied throughout. To probe informational sufficiency,
additional reconstruction models were trained on systematically restricted
subsets of two-body and three-body marginals.  

Separate feed-forward neural networks were developed for the
classification and reconstruction tasks. The classification networks predict
whether a given collection of reduced density matrices uniquely determines the
global state, whereas the reconstruction networks learn the mapping from
reduced density matrices to the corresponding global four-qubit density matrix.
The network architectures and training parameters were optimized independently
for each task to achieve the best predictive performance.

Architectural details, feature dimensions, and hyperparameters are summarized
in the Supplementary Material.
%%%%%%%%%%%%%%%%%%%%%%%%%%%%%%%%%%%%%%%%%%%%%%
\subsection{Experimental Benchmarking} We benchmark our marginal-based neural
framework on a four-qubit nuclear magnetic resonance (NMR) quantum processor,
using the four ${}^{13}$C spins of labeled trans-crotonic acid as qubits.
Entangled four-qubit states are prepared, their global density matrices
reconstructed by quantum state tomography, and the corresponding two- and
three-party marginals extracted.  These experimentally measured marginals are
then supplied to the trained networks for classification and reconstruction.
The NMR platform provides a controlled setting with intrinsic phase-damping
decoherence, enabling a direct test of robustness against realistic
experimental noise.
	
In the weak coupling regime, the rotating-frame internal Hamiltonian
is~\cite{OLIVEIRA}: \begin{equation} H = - \sum_{i=1}^{4} (\omega_i -
\omega_{\text{rf}}) I_i^z + \sum_{i<j}^{4} 2 \pi J_{ij} I_i^z I_j^z,
\end{equation} The system is initialized into a four-qubit pseudopure state
using spatial averaging~\cite{CORY1998}, achieving a fidelity of $0.9903$.  All
control sequences are decomposed into single-qubit rotations and $J$-coupling
evolutions, with two-qubit gates implemented via GRAPE-optimized
pulses~\cite{KHANEJA2005,Bhole2020}, robust to RF inhomogeneity and designed to
minimize total circuit duration ($\sim25$-$106$~ms).
	
Full four-qubit quantum state tomography was performed using constrained convex
optimization~\cite{Gaikwad2021}. The reconstructed experimental density
matrices achieve fidelities $\gtrsim 0.88$ for representative entangled states
and $\gtrsim0.94$ for random states.  Reduced density matrices are obtained by
partial trace and encoded using the same preprocessing pipelines as in
training. Reconstructability is evaluated with the same SDP thresholds used in
dataset labeling, and reconstruction quality is quantified by the fidelity
between experimental states and neural predictions.  All experimental
parameters, pulse sequences, and tomography details are provided in the
Supplementary Material.

%%%%%%%%%%%%%%%%%%%%%%%%%%%%%%%%%%%%%%%%%%%%%%%%%%%%%%%%%%%%
\section{Conclusions} 
\label{concl} 
We undertook a combined numerical and
neural-network study of the quantum marginal problem in four-qubit systems,
showing that marginal reconstructability is structured by entanglement class,
marginal order, and noise. Using semidefinite programming, we mapped the
certified reconstructability landscape and identified which families of states
are uniquely fixed by their reduced density matrices, demonstrating that
uniqueness is not generic but tied to genuinely multipartite correlations.
Next we showed that neural networks can learn this structure directly from
reduced data: they can infer reconstructability and when uniqueness holds, can
reconstruct global four-qubit density matrices from marginals alone with high
fidelity. We validated this framework on a four-qubit NMR quantum processor,
demonstrating that marginal-based neural reconstruction can be a realistic
state-characterization tool.  

By systematically restricting the available marginals, we identified
information thresholds and minimal sufficient sets, revealing that global
quantum structure can be recovered from surprisingly small subsets of reduced
data and pointing toward reduced-measurement strategies for quantum tomography
on noisy platforms.

Finally, we investigated the influence of noise on neural inference and
found that robustness depends strongly on the training strategy. Models trained
on datasets incorporating realistic noise generalized reliably across different
noise conditions, whereas models trained solely on ideal states showed a marked
loss of performance under experimental imperfections. These results demonstrate
that noise-aware learning is essential for practical quantum-state
reconstruction from reduced data.

%%%%%%%%%%%%%%%%%%%%%%%%%%%%%%%%%%%%%%%%%%%%%%%%%%%%%%%%%%%
\begin{acknowledgments} 
All the experiments were performed on a Bruker Avance-III 600 MHz FT-NMR
spectrometer at the NMR Research Facility of IISER Mohali.  M.~K.
acknowledges the DST-INSPIRE fellowship from the Department of Science
and Technology India for financial support.
\end{acknowledgments}
%\bibliographystyle{apsrev4-1} 
%\bibliography{ann-4entang} 


\end{document}